\documentclass[conference,10pt]{IEEEtran}
\usepackage[utf8]{inputenc}
\usepackage{graphicx}    

\usepackage{pifont}
\newcommand{\cmark}{\ding{51}}%
\newcommand{\xmark}{\ding{55}}%

\usepackage{amsmath}
\usepackage{chngcntr}
\usepackage{array}
\usepackage{epsf}
\usepackage{multirow}
\usepackage{rotating}
\usepackage{caption}
\usepackage{epsfig}
\usepackage{flushend}
\usepackage{subcaption}
\usepackage{epstopdf}
\usepackage{algorithm}
\usepackage{algorithmic}
\usepackage{xcolor}
\usepackage{cite}
\usepackage{float}

\usepackage{enumitem}

\RequirePackage[normalem]{ulem} 
\RequirePackage{color}\definecolor{RED}{rgb}{1,0,0}\definecolor{BLUE}{rgb}{0,0,1} 

\title{Channel Selection Scheme for Cooperative Routing Protocols in Cognitive Radio Networks\vspace{-2ex}}

\author{
  \IEEEauthorblockN{\makebox[.33\linewidth]{Arsany Guirguis}}
  \IEEEauthorblockA{
   Comp. and Sys. Eng. Dept.\\
   Alexandria University, Egypt\\
    arsany.guirguis@alexu.edu.eg
  \vspace{-2ex}}
\and
 \IEEEauthorblockN{\makebox[.33\linewidth]{Mustafa El-Nainay}}
  \IEEEauthorblockA{
   Comp. and Sys. Eng. Dept.\\
   Alexandria University, Egypt\\
    ymustafa@alexu.edu.eg
  \vspace{-2ex}}

}

\def \sys {\textit{CSCR}}
\begin{document}
\maketitle

\begin{abstract}
Cognitive radio networks (CRNs) propose a smart solution for spectrum usage inefficiency. Routing protocols for CRNs follow different criteria to choose the best route to the destination and to avoid the interference with primary users. Some protocols use cooperative communication techniques to achieve the full coexistence between secondary users (SUs) and primary users (PUs). Although using such cross-layer techniques have a great impact on the quality of the chosen routes, the existing work do not choose the best channel to send the data over. Thus, the available spectrum is not utilized efficiently. In this work, we propose \sys{}, a channel selection scheme for cooperation-based routing protocols in CRNs. The proposed scheme increases the spectrum utilization through integrating the channels selection in the route discovery phase of the cooperation-based routing protocols. The best channels, that are less congested with primary users and that lead to minimum switching overhead, are chosen while constructing the cooperative group. Evaluating \sys{} via NS2 simulations shows that it outperforms its counterparts  in terms of goodput, end-to-end delay, and packet delivery ratio. The proposed scheme can enhance the network goodput, in some cases, by more than 150\%, as compared to other related protocols.
\end{abstract}

\section{Introduction}

Cognitive Radio Networks (CRNs) appeared as a promising solution for the spectrum underutilization problem. In these networks, the primary users (PUs) are the only licensed users to use the spectrum. However, secondary users (SUs) are able to use it too, but under the condition of not interfering with the PUs. Several routing protocols were designed to construct routes between SUs in CRNs \cite{crn_metric_survey}, each has its own criteria. In this paper, we are extending Undercover \cite{undercover} which utilizes the cooperative communication techniques in the routing process. Using these cross-layer techniques, the signal quality can be enhanced in the direction of the receiver SU (cooperative diversity), while nulling the transmissions in the directions of the PUs (cooperative beamforming). Although this idea allows the full coexistence between SUs and PUs, the existing work ignore choosing the proper channel to transmit signals on, and this leads to an inefficient use of the spectrum.





In this paper, we propose \sys{}: a Channel Selection scheme for Cooperative Routing protocols in CRNs. This scheme aims at choosing dynamically the best channel to use at each relay along the route. It avoids choosing the channels with high PUs activities while having the least possible channel switching delay. This strategy enhances the network metrics in terms of achievable capacity, end-to-end delay, and packet delivery ratio. Although the core idea has been used previously in the CRNs literature, the task of selecting the best channel for a cooperative group (during routing) was not discussed before. Thus, this is the first work that investigates this problem and integrates the channel selection with a cooperative \emph{routing} protocol.


\begin{figure}[!t]
\centering
	\begin{subfigure}[t]{0.15\textwidth}
	\centering
    \includegraphics[width=1in]{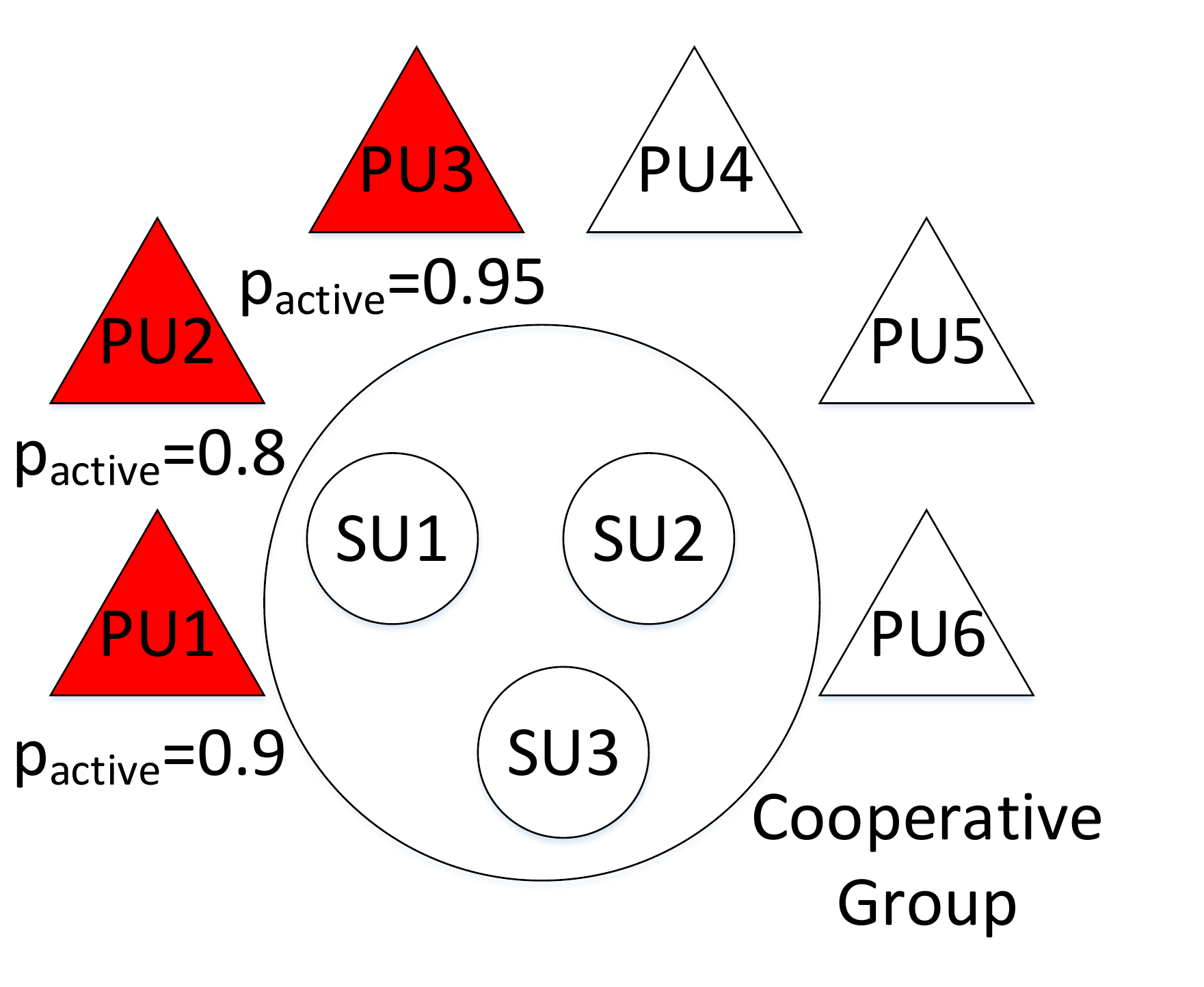}
	\caption{Channel 1}
	\label{fig::channel1}
	\end{subfigure}
	\begin{subfigure}[t]{0.15\textwidth}
	\centering
    \includegraphics[width=1in]{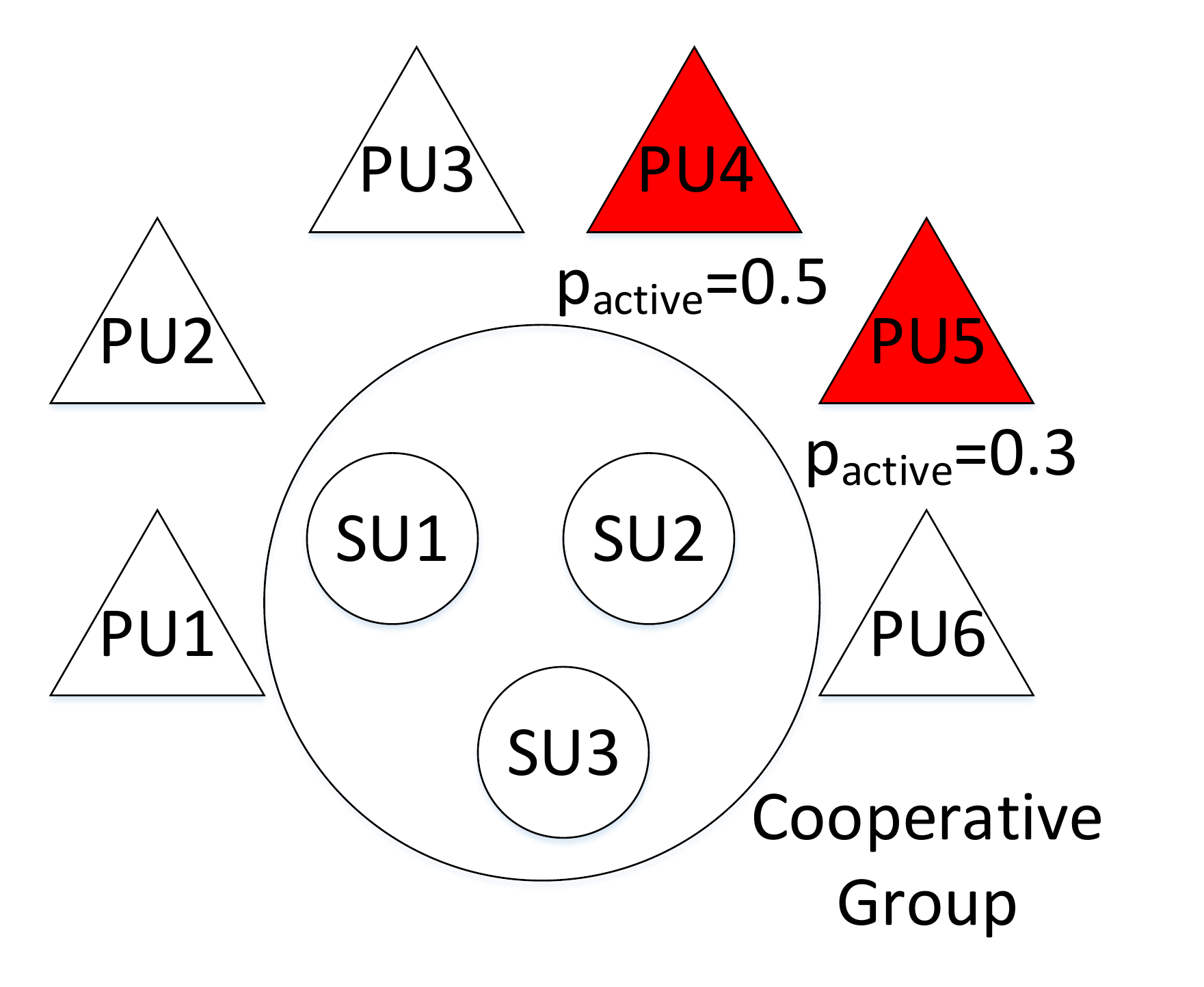}
	\caption{Channel 3}
	\label{fig::channel3}
	\end{subfigure}
	\begin{subfigure}[t]{0.15\textwidth}
	\centering
    \includegraphics[width=1in]{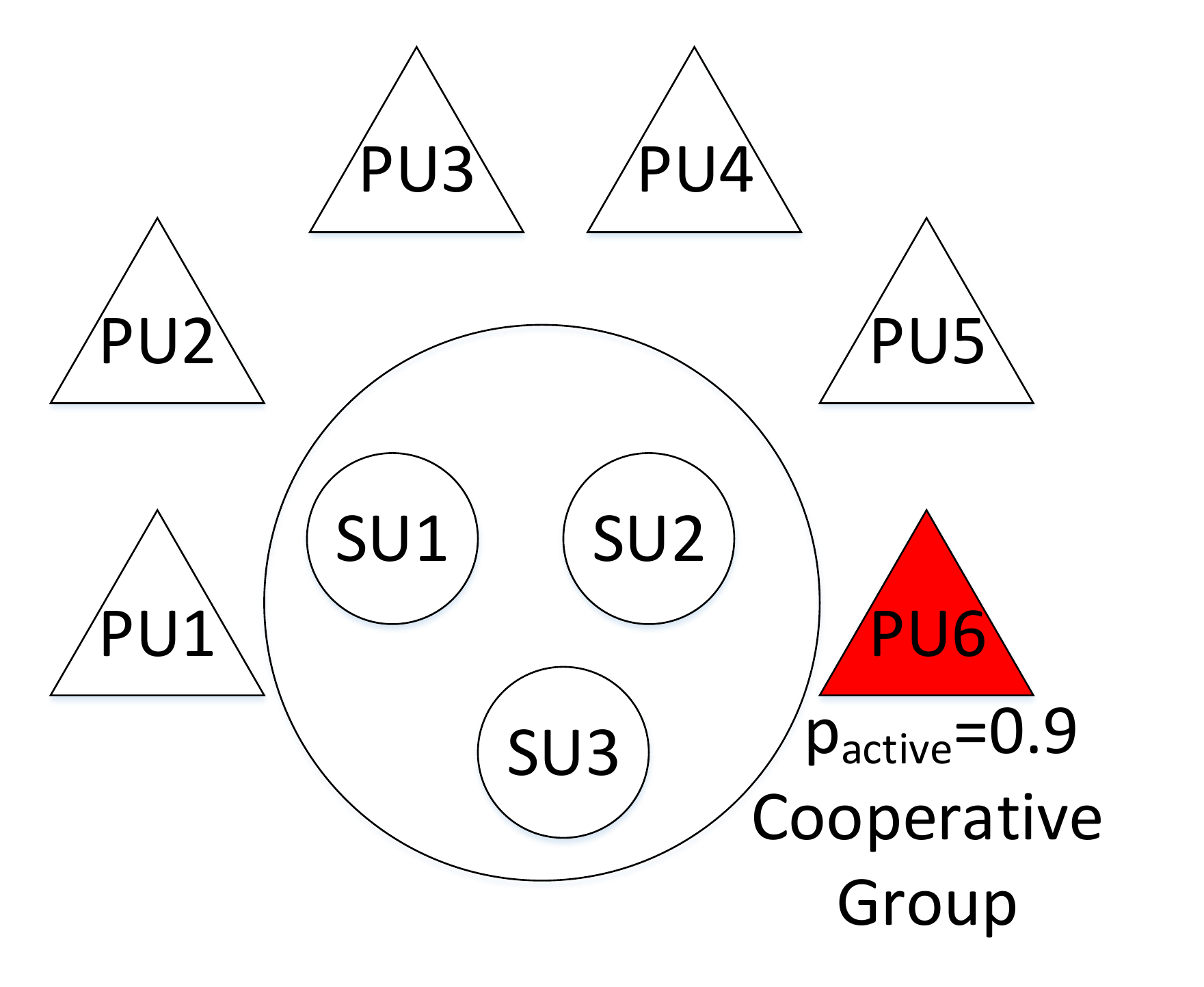}
	\caption{Channel 10}
	\label{fig::channel10}
	\end{subfigure}
\caption{Example on the effect of choosing the optimal channel to send the data over. Note that $p_{\text{active}}$ represents the activity probability of a primary user on a particular channel.}
\label{fig::example}
\vspace{-5ex}
\end{figure}


Our scheme is motivated by the example in Figure~\ref{fig::example} in which the advantage of choosing the appropriate channel, to send the data over, is shown. Consider having a cooperative group of three SUs that are required to select a channel (channel 1 is the default) to send some data through. Sending on the default channel (channel 1) is not the best choice due to the high number and activity of the surrounding PUs (Figure~\ref{fig::channel1}). Also, sending on channel 10 (Figure~\ref{fig::channel10}) is not the best choice too, since switching from the default channel to channel 10 will take a long time, as the latter channel is the farthest one from the default channel\cite{gozupek2013spectrum}. Thus, the best channel is channel 3 (Figure~\ref{fig::channel3}) because, it achieves the compromise between avoiding highly active PUs and minimizing the channel switching overhead.

The proposed scheme has been evaluated using NS2 \cite{NS2}, and its performance is compared to two other CRN protocols.  Results show that \sys{} can enhance the network goodput, in some cases, by more than 150\%. In addition, it is shown that \sys{} always experiences a higher data delivery ratio, with the least end-to-end delay, compared to its counterparts.

The rest of the paper is organized as follows: Section \ref{sec::related_work} presents the related work. Then, our system model is described in Section \ref{sec:sysModel}. The proposed channel selection scheme is given in Section \ref{sec::impdetails}. Section \ref{sec::eval} evaluates \sys{}  and Section \ref{sec::conc} concludes the paper and gives directions for future work.
\section{Related Work}
\label{sec::related_work}
In this section, we present the previously published work in using cooperation-based protocols and channel selection schemes in CRNs.

\subsection{Cooperation-based Protocols}
In the wireless medium, signals can constructively or destructively collide with each other and hence their characteristics are changed. Researchers used this idea to intentionally change the signals behavior for different goals. This is done by transmitting some signals cooperatively at the same time by a group of users \cite{goldsmith2005wireless,tse2005fundamentals}. This idea helped in enhancing throughput \cite{lakshmanan2009diversity}, saving energy \cite{khandani2005cooperative}, enabling simultaneous non-interfering signals \cite{collier1991transmission}, and reshaping the transmission beam \cite{van1988beamforming}.


After its success in conventional wireless networks\cite{khandani2007cooperative}, cooperative communication techniques were used extensively in CRNs\cite{zhang2009cooperative,ding2010distributed}. For instance,
\emph{cooperative diversity} was used to transmit data with higher capacity to
maximize throughput \cite{ding2010distributed} and to enhance the end-to-end
throughput\cite{sheu2012cooperative}. On the other hand, DZP \cite{karmoose2013dead} uses \emph{cooperative beamforming} in implementing a route maintenance scheme. It allows using the same already-established routes even if a PU becomes active. Thus, it alleviates the overhead of re-establishing new sub-optimal routes. This is done through transmitting signals by a cooperative pair (a group of two SUs) in a way that nulls out the transmission at the active PU. Along the same line, Undercover\cite{undercover} uses cooperative beamforming in the \emph{routing} process. At each hop of the route, a cooperative group of SUs is chosen to send the data simultaneously. It also utilizes cooperative diversity to enhance the signal quality in the SU receiver direction.

\subsection{Channel Selection}
Routing protocols in multi-channel \emph{wireless} networks take two main decisions: (1) choosing a route to the destination and (2) choosing channels used for transmission at each hop on this route. Thus, channel selection is considered an integral part of the routing process in such networks \cite{nasipuri2000multichannel, mumey2012routing}. The main goals of the channel selection, in this case, are to achieve the maximum utilization of the spectrum and to decrease the interference between active flows as much as possible.

Specifically, in the context of CRNs, Tragos et al.\cite{tragos2013spectrum} and Saleem et al.\cite{saleem2015routing} presented surveys that give some approaches of utilizing the available channels to achieve different goals. In some publications \cite{joshi2012joint,salameh2011throughput}, the channel assignment problem has been formulated as an optimization problem with different objectives. Joshi et al., in \cite{joshi2012joint}, aimed at using the minimum power that can be used to transmit data successfully without being considered noise. However, Salameh et al., in \cite{salameh2011throughput}, sought to maximize the achievable capacity. A brute force search for the best channel, according to different metrics, has been proposed in other publications \cite{lee2011spectrum,kim2010urban}. Lee et al., in \cite{lee2011spectrum}, aimed at choosing the ones that minimize the end-to-end delay for all connections. Kim et al., in \cite{kim2010urban}, chose the channel that is not congested with highly-active PUs. Along the same line, Li et al., in \cite{li2010enhancing}, proposed a heuristic algorithm that seeks to have reliable and robust paths, in terms of avoiding PUs, for the secondary network. However, some important parameters were not taken into consideration in the formulations proposed in these publications. For example, Salameh et al., in \cite{salameh2011throughput}, did not account for the PUs activities in their proposed solution. Also, Kim et al., in \cite{kim2010urban}, ignored the switching delay and the SUs flows interference on each other. Finally,  Li et al., in \cite{li2010enhancing}, did not consider maximizing the achievable capacity.

In this paper, we consider solving the channel selection problem for a cooperation-based routing protocol. The main additional requirement in the cooperative case, than the default case of sending using only one node, is that: all group members should operate on the \emph{same channel}. This allows the transmitted signals to null each other at PUs directions. The channel selection algorithm aims at choosing the channel that achieves (1) the best achievable capacity for SUs flows, (2) the least interference between SUs flows and each other, (3) the least impact on surrounding PUs, and (4) the minimum channel switching delay. We combine all these requirements in a routing metric which is used to determine the best group to relay the data, along with the best operating channel.

\section{System Model} \label{sec:sysModel}



Our protocol is designed to work in a CRN which consists of two types of users: PUs and SUs. PUs have the license to use the spectrum according to their data delivery requirements. Moreover, we assume that SUs can detect and sense the activity of the surrounding PUs. Primary network is supposed to adopt an overlay transmission policy which means that SUs can transmit data in two cases only: (1) surrounding PUs are not currently active, or (2) SUs are able to use cooperative beamforming so that their transmissions do not interfere with the PUs. For the wireless channels at all nodes, the slow fading multipath model is assumed, in which the channel coefficients are constant over some specified time period. Reliable estimates for the channel coefficients, between SUs and each other and between SUs and PUs, can be done through the channel estimation techniques on the preambles of transmitted packets between different nodes \cite{de1997maximum}. Through utilizing these channel coefficients, an accurate beamforming can be implemented. SUs are assumed to transmit control packets over a Common Control Channel (CCC) such as the 2.45GHz ISM band.

While Undercover\cite{undercover} assumed that all nodes have only one channel to send the data over, we released this assumption in \sys{}. All nodes in the CRN (SUs and PUs) can operate on multiple channels. The sending channel may differ from the receiving channel, and each node is able to switch to any of these channels at any time. Channel switching delay is assumed to depend on the difference between frequencies of the current and the target channels \cite{habak2013location,gozupek2013spectrum}. Channels are assumed to be non-overlapping which means that sending data on one channel does not interfere with the transmitted data on other channels.

\section{Proposed Channel Selection Scheme for Cooperative Routing}\label{sec::impdetails}
In this section, we give an overview for the proposed channel selection scheme for cooperative routing in CRNs. First, we present the goals which we considered while designing \sys{} and then we show how we achieve these goals through the model we propose. Then, we give details of the implementation and the flow of the proposed channel selection scheme.

\subsection{Design Goals}
We have designed \sys{} to achieve some goals which ensure feasibility, usability, and efficiency of the proposed protocol. First, \sys{} is designed in a decentralized approach, where each group can \emph{independently} choose the channel it will send the data over. Second, as a protocol planned to operate in CRNs, the PUs activities are considered while choosing the operating channel. Taking this into consideration gives more reliable routes that interfere less with the surrounding PUs. Also, we take care of the changing environment of CRNs which includes the periodic changes in the number of active flows between all nodes (including SUs and PUs), the PUs activities, and the channels' conditions.

Since we assume a multi-channel environment, the channel switching delay is taken into consideration. We aim at minimizing this overhead as much as possible, given that all participating nodes in the group should operate on the same channel while sending data. Although switching of all group members to one channel, to send data, is costly, one may not have another choice if all channels are occupied by active PUs. Another important parameter we take into consideration is the operating channels of the other active flows passing by the nodes participating in the chosen group. Switching the sending channel of one of the group members while one of its flows is active may cause preemption of this flow and loss of data. Thus, we aim at choosing the channels that avoid causing this preemption. Finally, we take care of the same goals of Undercover\cite{undercover} which are: increasing the achievable capacity and minimizing the interference with other SUs flows.

\subsection{Routing Metric} \label{sec:routingmetric}
In this section, we present how the mentioned goals are achieved through the proposed routing metric. First of all, the group construction method is inherited from Undercover protocol\cite{undercover}. The channel selection algorithm is added to the group construction phase, so that a group can choose the best channel to work on. This removes the overhead of choosing a channel while sending the data and also, decreases the possibility of losing data due to the potential interruption/contention caused by other active flows. Based on this, each group in the network (even on the same flow path) chooses independently and periodically the optimal channel for data transmission. This periodic decision is done so that the algorithm can adapt to the changing network environment.



We define the routing metric\footnote{We have inherited the first three terms from Undercover metric\cite{undercover}.} that node $i$ (which has some data to send) or a relay can achieve, with the help of a chosen cooperative group, while sending data to node $j$ by:
\begin{equation}
\label{eqn:metric}
LC_{ij} = \dfrac{\hat{C}_{ij}}{\big(N_n + \beta (N_f - N_n)\big) \times p_{\text{pu}} \times T_{\text{switch}}}.
\end{equation}
\noindent Where:
\begin{itemize}[noitemsep,topsep=0pt,parsep=0pt,partopsep=0pt]
\item $\hat{C}_{ij}$ is the maximum achievable capacity between nodes $i$ and $j$ among all checked groups. This term depends on the available bandwidth, maximum power can be achieved by the node, and the channel coefficients between group members and the receiver node.
\item $N_n$ and $N_f$ represent the interference caused from the outgoing flows to the group members and due to the group construction on other flows, respectively. Specifically, $N_n$ is the number of direct neighbors, of all group members, which carry active flows, and $N_f$ is the number of flows that are in the interference range of the coopeartive group.
\item $\beta$ is a user-defined design parameter to alter the possible altruistic behavior of the cooperative group.
\item $p_{\text{pu}}$ defines the probability of at least one of the surrounding PUs to be active, on the checked channel,  within some specified time period $\tau$. According to our two-state ON-OFF birth-death PUs model, this  is given by \cite{habak2013location}:
\begin{equation}
p_{\text{pu}} = 1 - e^{- \tau \sum_{i=1}^{n_{pu}} \mu_i}.
\end{equation}
\noindent where $n_{pu}$ is the number of surrounding PUs that are active on the considered channel and $\mu_i$ is the parameter of the exponential distribution of the OFF period of $PU_i$. Thus, this term traces the surrounding PUs effect.
\item $T_{\text{switch}}$ represents the delay cost that results from switching of all group members to the selected channel. This is given by:
\begin{equation}
T_{\text{switch}} = \underset{m \in g}{\max} \, d_{\text{chan}_m} \times c.
\end{equation}
\noindent where, $g$ is the set of cooperative group members, $d_{\text{chan}_m}$ represents the distance between the current channel of node $m$ and the target channel, and $c$ is a constant that reflects the cost of switching between two consecutive channels. Since all group members switch their channels simultaneously to the target one, the total switching cost is the delay that results from the node that has the farthest channel from the target one.
\end{itemize}

Finally, it is important to note that this metric is calculated for each available valid channel while constructing a group. A channel is considered to be valid in two cases only: (1) there is no other active flows on this node or (2) all flows passing through this node are transmitted using this channel. These conditions ensure that there will be no channel switching while a node is transmitting data. This gives the protocol more reliability in delivering the data from the source to the destination. If no channel with the specified criteria is found, the algorithm chooses to use the channel that costs the minimum switching delay. 

\subsection{Information Exchange}
In order to implement the channel selection task, each node should know a set of information which may change from time to time. This set of information includes:
\begin{enumerate}
\item The node direct neighbors, their available channels, and all channels coefficients between them and the node.
\item All flows passing by the node neighbors and the channels they are working on.
\item All PUs that are sensed by the direct neighbors along with their activity probabilities, activity channels, and the channels coefficients between them and the node neighbors.
\end{enumerate}
These pieces of information are communicated between the nodes through periodic ``Hello" packets.
In addition, each node takes some parameters into account while choosing a channel to work on including (1) IDs of the available channels, (2) the currently occupied channels by other flows, and (3) the current sending channel.

 \begin{figure}[!t]
 \centering
  \includegraphics[width = 0.4\textwidth]{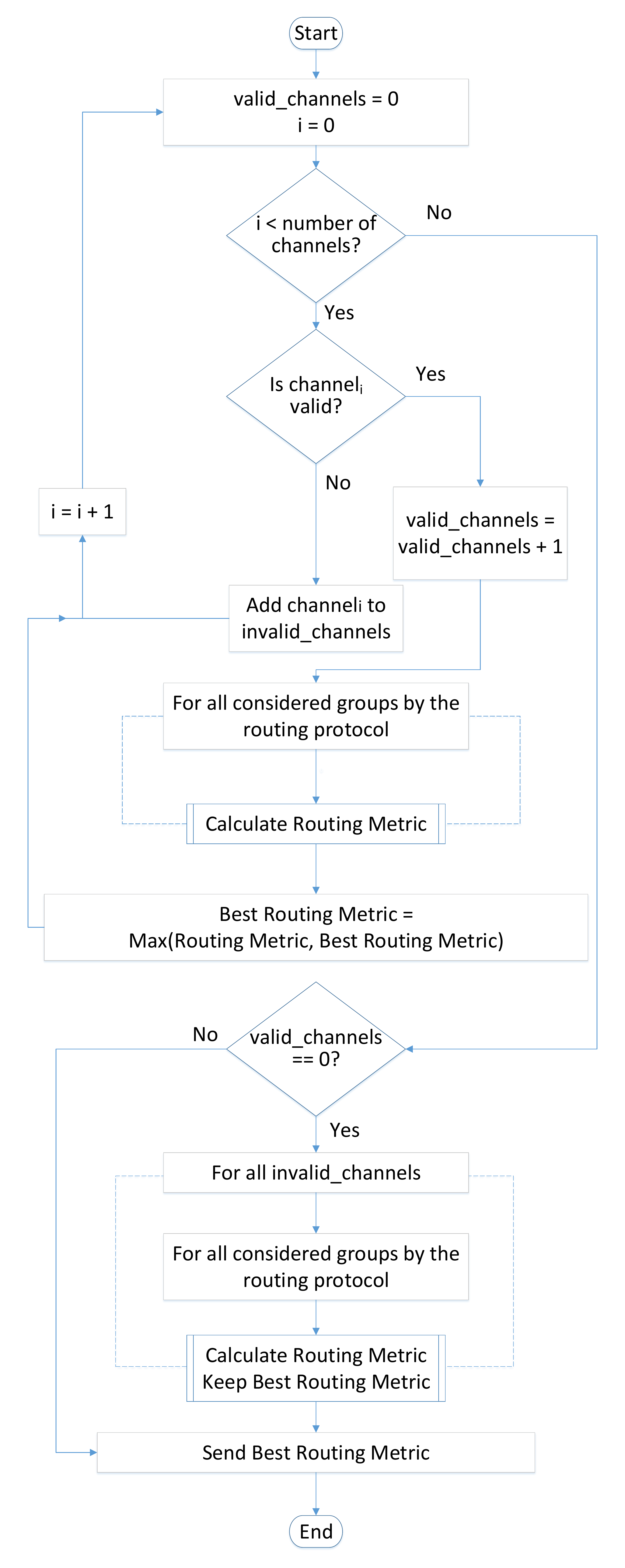}
  \caption{Flowchart of the channel selection algorithm applied by the node required to construct the cooperative group.}
  \label{fig::flowchart}
\vspace{-0.35in}
 \end{figure}

\subsection{Channel Selection Algorithm Flow}

Figure~\ref{fig::flowchart} gives the flowchart of the entire algorithm. The main part of the protocol (which searches for the best channel to use) works in the group construction phase. In this phase, all channels are checked for their feasibility. If the channel is infeasible, it is added to the invalid channels set and the next channel is checked. However if the channel is valid, different cooperative groups are considered by the routing protocol\footnote{According to \cite{undercover}, the possibility of using one node in sending data, instead of using a cooperative group, is also checked at this stage. This ensures that the protocol will not use beamforming unless it is really useful.}. For each group, the routing metric is computed according to Equation (\ref{eqn:metric}) and is kept with the node, if it is better than the best metric the node has got. At the end, the best value of routing metric is considered. However, if one cannot find any valid channel, the invalid channels are checked along with the groups considered by the routing protocol. In this case, the channel that achieves the minimum switching delay is chosen. Finally, this node can send to the source node the best routing metric which represents a chosen group for sending data, along with a channel that is recommended to work on.

\section{Performance Evaluation} \label{sec::eval}

In this section, the performance of \sys{} is evaluated in different network configurations. First, the simulation setup and the used parameters are presented. Then, the metrics used for the evaluation process are defined. Finally, we present the results of our experiments.
\subsection{Simulation Setup}
We used a cognitive extension of NS2 \cite{NS2},\cite{crextension} for simulation. Table \ref{fig::simulation_parameters_summary} introduces the simulation parameters used in the evaluation. We model the PUs activities as an ON-OFF process. The means of the exponentially-distributed active and inactive periods are randomly chosen with the activity percentage shown in Table \ref{fig::simulation_parameters_summary}. We follow the same assumptions stated in Section \ref{sec:sysModel}. The used MAC layer protocol is IEEE 802.11 protocol. The source and the destination of each connection are selected randomly. We compare \sys{} against two other protocols which are LAUNCH \cite{habak2013location} and Undercover \cite{undercover}. LAUNCH is a location-aided \emph{channel-aware} routing protocol that is designed to work in CRNs. However, it does not use beamforming in the routing process. On the other hand, Undercover uses \emph{beamforming} in routing, but it never knows about the existence of channels nor how to benefit from them. We have chosen these two protocols to show the advantage of the main properties of our integrated channel selection scheme with the cooperative routing protocol. Table \ref{table:prop} summarizes the target properties of the three protocols.

\begin{table}[!t]
    \centering
    \caption{Experiments parameters.}
    \begin{tabular}{|l |l |l|}
    \hline
    Parameter & Value range & Nominal Value\\
    \hline
    Number of PUs & 0 - 16 & 8\\
    Number of SUs & 10 - 30 & 25\\
    SU transmission range (m) & 125 & 125\\
    PU transmission range (m) & 140 & 140\\
    Number of connections & 1 - 16 & 8\\
    Effective bandwidth (Mbps) & 1.5 & 1.5 \\
    Packet Size (byte) & 512 & 512\\
    Number of available channels per node &   3 - 9 & 5 \\
    PU Activity(\%) &   20 - 80 & 40 \\
    Data Rate Per Source (kbps) & 100 & 100 \\
    Deployment Area Side Length (m) & 250 & 250 \\
    \hline
    \end{tabular}
    \label{fig::simulation_parameters_summary}
\end{table}

\begin{table}[]
\centering
\caption{Properties of the compared routing protocols.}
\label{table:prop}
\begin{tabular}{|l|c|c|}
\hline
Protocol   & Channel Aware & Beamforming \\ \hline
LAUNCH     & \cmark        & \xmark      \\ \hline
Undercover & \xmark        & \cmark      \\ \hline
\sys{}   & \cmark        & \cmark      \\ \hline
\end{tabular}
\end{table}

\subsection{Metrics}
We evaluate \sys{} using the following metrics:
\begin{enumerate}
\item Goodput: number of bits communicated successfully from the source to the destination per second.
\item Average end-to-end delay: average time taken by packets to reach the destination from the source.
\item Packet delivery ratio: percentage of packets reached the destination to the total number of packets generated by the source.
\item Average group size: average number of nodes participating in the cooperative communication in case of using \sys{}.
\item Routing overhead: number of transmitted control packets in the routing phase.
\end{enumerate}

\begin{figure*}[!t]
\centering
	\begin{subfigure}[t]{0.24\textwidth}
	\centering
    \includegraphics[width=1.8in]{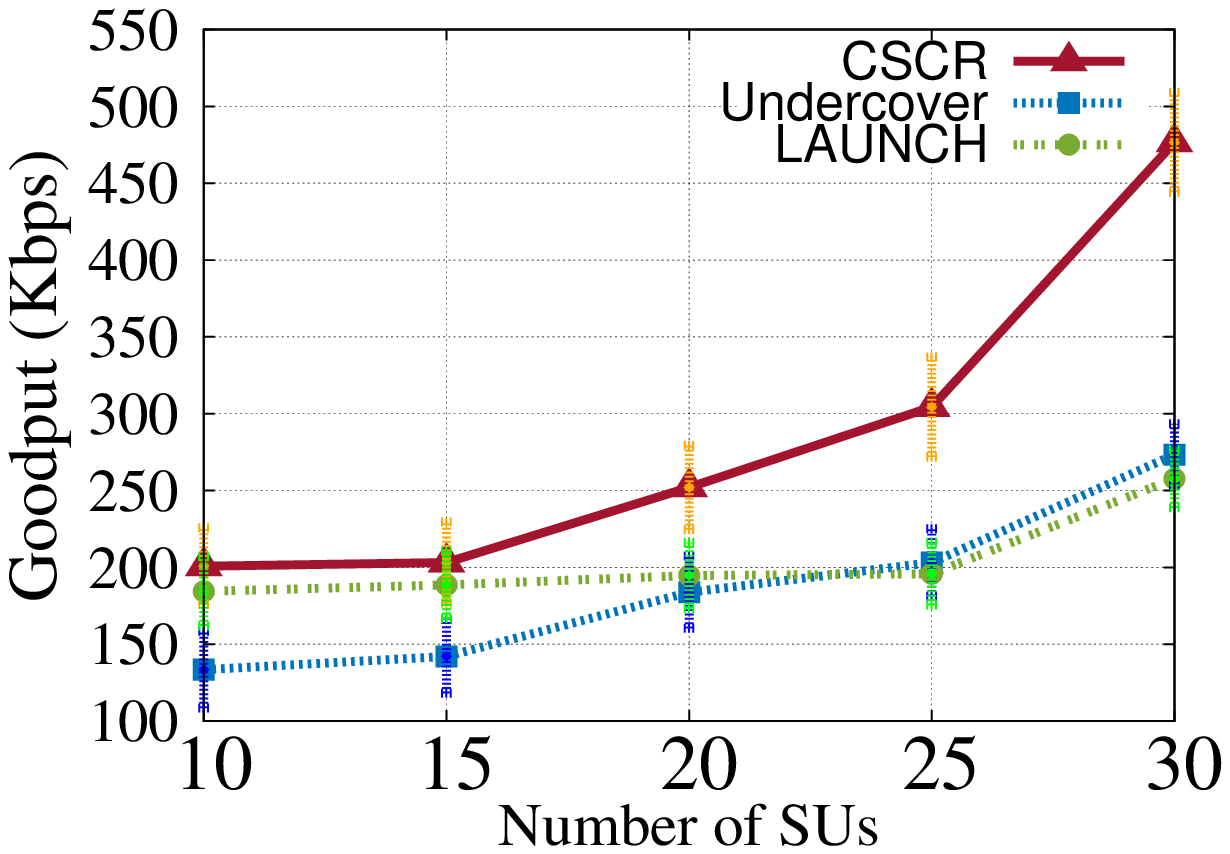}
	\caption{Goodput}
	\label{fig::nsusthrough}
	\end{subfigure}
	\begin{subfigure}[t]{0.24\textwidth}
	\centering
    \includegraphics[width=1.8in]{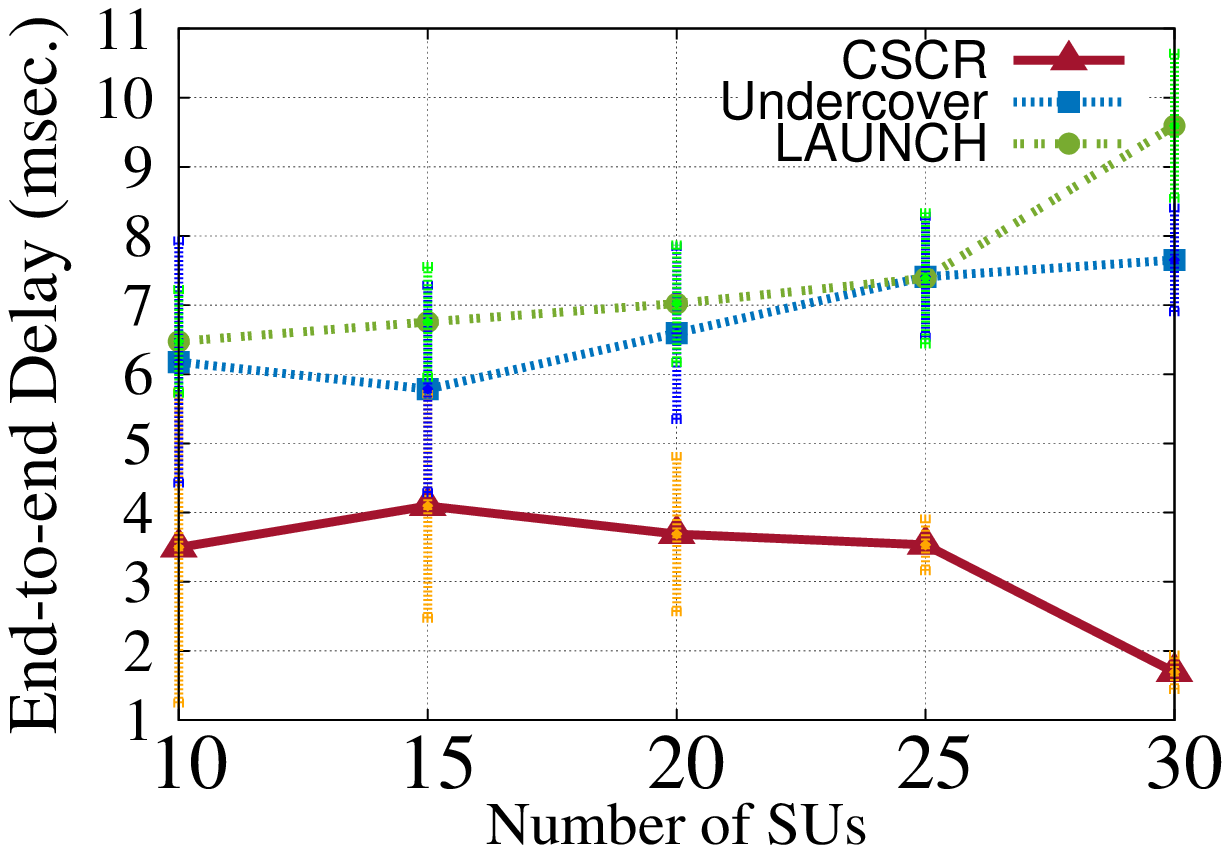}
	\caption{Average End-to-end Delay}
	\label{fig::nsusdelay}
	\end{subfigure}
	\begin{subfigure}[t]{0.24\textwidth}
	\centering
    \includegraphics[width=1.8in]{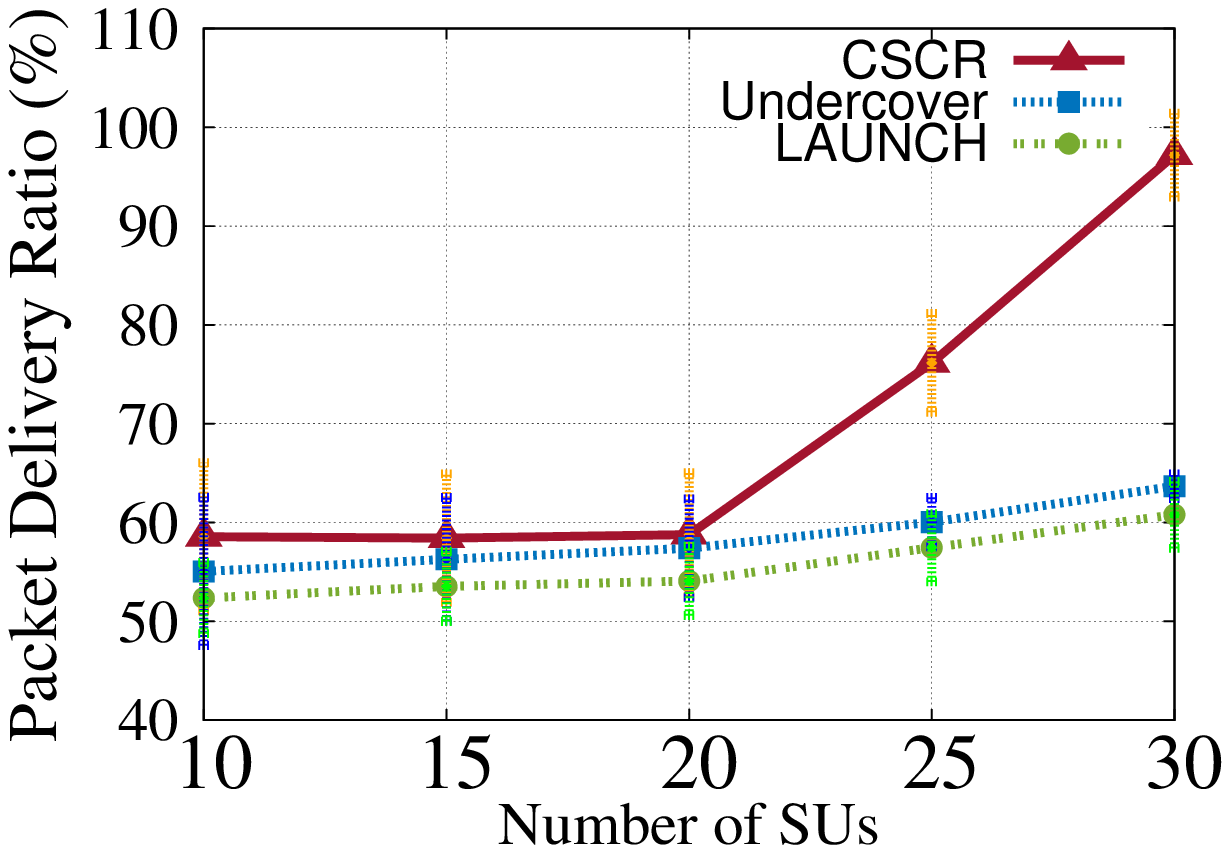}
	\caption{Packet Delivery Ratio}
	\label{fig::nsuspdf}
	\end{subfigure}
	\begin{subfigure}[t]{0.24\textwidth}
	\centering
    \includegraphics[width=1.8in]{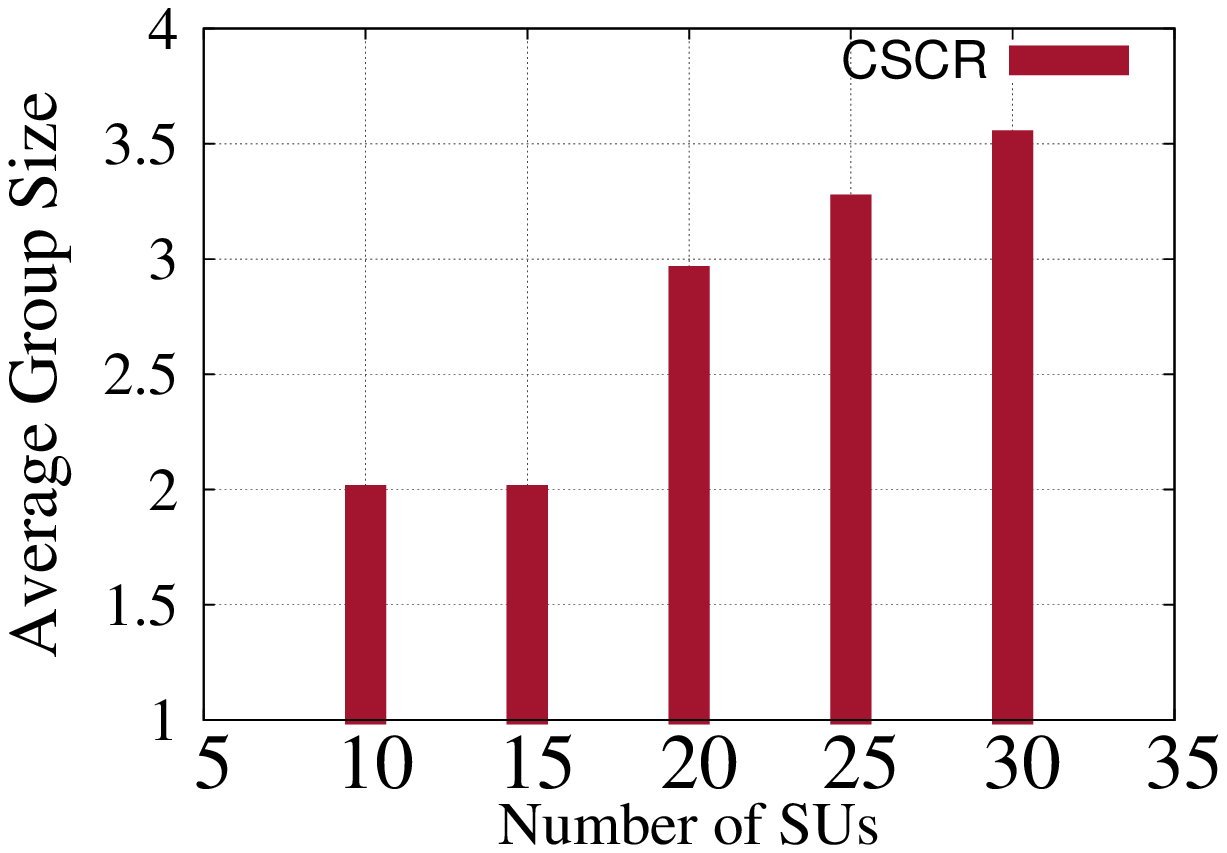}
	\caption{Average Group Size}
	\label{fig::nsusgpsize}
	\end{subfigure}
\caption{Effect of changing number of SUs on network performance.}
\label{fig::nsus}
\vspace{-0.35in}
\end{figure*}

\subsection{Experimental Results}
\subsubsection{Changing Number of SUs}
Figure \ref{fig::nsus} compares the performance of the three protocols while changing the number of SUs. Generally, it shows the advantage of \sys{} over both Undercover and LAUNCH protocols, in terms of the measured metrics. We can derive some key conclusions from Figure \ref{fig::nsusthrough}. First, we can see that the goodput is enhanced as the number of SUs increases. This is due to the availability of new opportunities and routes as the SUs' density increases. On the same line, we can see that the goodput increase for \sys{} is even higher than that of other protocols. We can reason this back to the ability of \sys{} to construct cooperative groups (as shown in Figure \ref{fig::nsusgpsize}) and use beamforming on the best available channels. In some experiments, groups of six nodes are attained and the average number of constructed groups exceeds $225$ groups per flow. Although Undercover uses cooperative groups, it achieves lower goodput than \sys{}. This happens since Undercover chooses a random channel to send the data on, which may not be the best channel. However, the advantage of using these groups can be observed at high SUs' density where there are many opportunities to construct groups. On the other hand, LAUNCH chooses the best channel to send on, but it lacks the advantage of using cooperative groups. To summarize, we can observe that the performance of Undercover and LAUNCH are near to each other and both of them achieve goodput lower than that of \sys{} in almost all configurations.

	\begin{figure}[!t]
	\centering
    \includegraphics[width=2.5in]{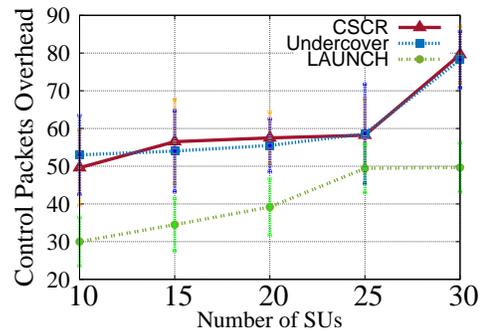}
	\caption{Control packets overhead of the three protocols.}
	\label{fig::nsusoverhead}
\vspace{-0.35in}
	\end{figure}

\begin{figure*}[!t]
\centering
	\begin{subfigure}[t]{0.32\textwidth}
	\centering
    \includegraphics[width=2.2in]{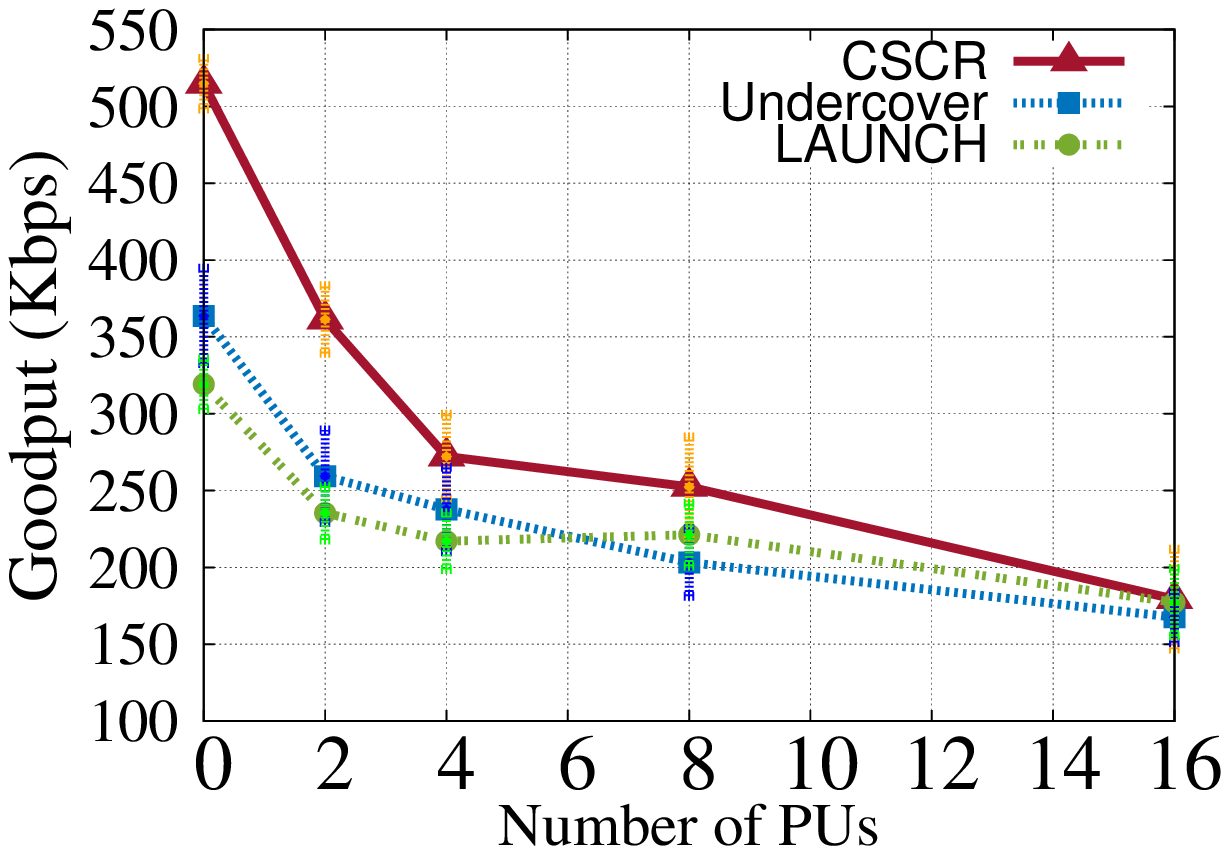}
	\caption{Goodput}
	\label{fig::npusthrough}
	\end{subfigure}
	\begin{subfigure}[t]{0.32\textwidth}
	\centering
    \includegraphics[width=2.2in]{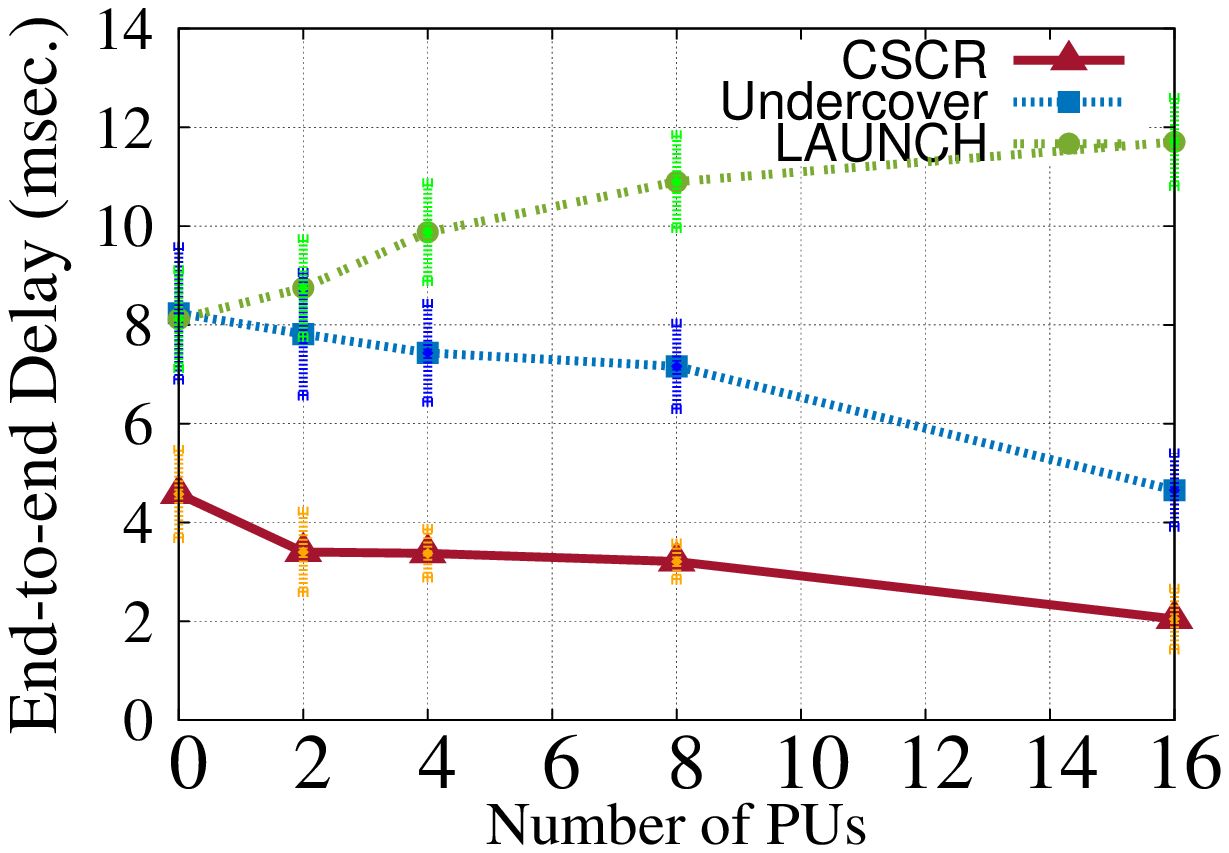}
	\caption{Average End-to-end Delay}
	\label{fig::npusdelay}
	\end{subfigure}
	\begin{subfigure}[t]{0.32\textwidth}
	\centering
    \includegraphics[width=2.2in]{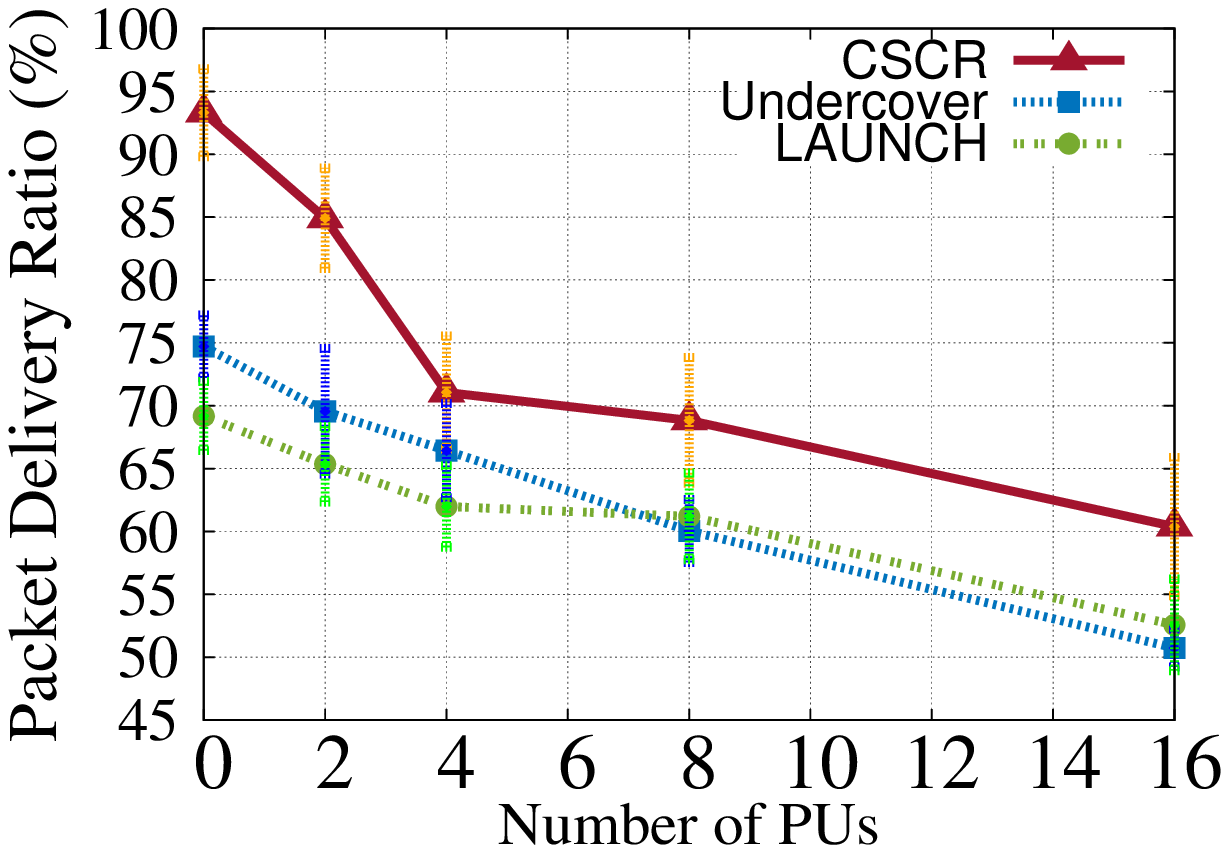}
	\caption{Packet Delivery Ratio}
	\label{fig::npuspdf}
	\end{subfigure}
\caption{Effect of changing number of PUs on network performance.}
\label{fig::npus}
\vspace{-0.3in}
\end{figure*}

Taking the discussion to Figure \ref{fig::nsusdelay}, we can see that the behaviors of the three protocols differ from each other as the number of SUs increases. We start at first with LAUNCH and Undercover protocols which have the same behavior. Although new and better routes can be attained when the number of SUs increases (as discussed in Figure \ref{fig::nsusthrough}), the average end-to-end delay of all packets increases too. This is attributed to the congestion at the MAC layer which results in increasing the number of transmissions between network nodes and increasing the queuing lengths and delays, increasing the total end-to-end delay. However, this is not the case for \sys{} where the average end-to-end delay decreases with the increase of SUs' density. This happens since \sys{} can choose the best channel to send on. This channel is the one that has the lowest PUs activities and the least interference with the existing flows. This choice leads to introducing new flows for data without interfering with either PUs or the existing SUs flows.

For this reason, we can observe that the packet delivery ratio is enhanced for \sys{} as shown in Figure \ref{fig::nsuspdf}. But, the enhancement of this ratio for other protocols is not large due to the increase of the noise interference with increasing the number of SUs. Yet, we can see that the delivery ratio of Undercover is always better than that of LAUNCH. This happens because the former protocol uses cooperative communication to overcome the activities of the PUs where LAUNCH cannot do more than choosing the best channel for sending data.

Finally, Figure \ref{fig::nsusoverhead} compares the control packets overhead of the three protocols with each other. Generally, as expected, the overhead increases with the number of SUs increase. We can notice that the number of control packets of \sys{} and Undercover is higher than that of LAUNCH; this is attributed to the extra packets sent by both former protocols to employ the beamforming. Comparing overhead of \sys{} and Undercover, we can observe that they are almost the same in all cases since, both of them transmit the same types of control packets. On the other hand, the difference between the control packets of \sys{} and LAUNCH is almost constant. Thus, this extra overhead of \sys{} can be outweighted by its higher throughput and packet delivery ratio performance, especially at higher number of SUs.

\subsubsection{Changing Number of PUs}
Figure \ref{fig::npus} shows the effect of changing the number of PUs on the performance of the three routing protocols. We can note that increasing the number of PUs leads to decreasing the goodput generally for the three protocols (Figure \ref{fig::npusthrough}). This happens due to the degradation of the packet delivery ratio as shown in Figure \ref{fig::npuspdf}. Both of these effects happen because of the increase of the PUs effect on the SUs links. However, in both figures, \sys{} outperforms Undercover and LAUNCH for two reasons. First, \sys{} is always able to choose the best channel to operate on. This gives the protocol the ability to avoid PUs' active channels. Second, \sys{} uses cooperative groups to even beat the areas where PUs are there and active. The final note we can draw from these two figures is that Undercover performs better than LAUNCH at low number of PUs. This happens since it can construct cooperative groups to overcome the PUs effect. However, in the case of the existence of high number of PUs, the effect of the improper choice of channels appears and Undercover performance becomes worse than LAUNCH.

Figure \ref{fig::npusdelay} tells us that the average end-to-end delays for both \sys{} and Undercover decrease with the increase in the number of PUs. This is due to the decrease in the number of transmissions and congestion as discussed in Figure \ref{fig::nsus}. Yet, only LAUNCH delay increases with the increase of the PUs density. This happens since LAUNCH adopts the interweave model in which SUs are not allowed to send any data if the surrounding PUs are active \cite{habak2013location}. This forces SUs on the route to wait for a long time before sending the data, keeping in mind the high activity of all PUs.

\begin{figure*}[!t]
\centering
	\begin{subfigure}[t]{0.32\textwidth}
	\centering
    \includegraphics[width=2.2in]{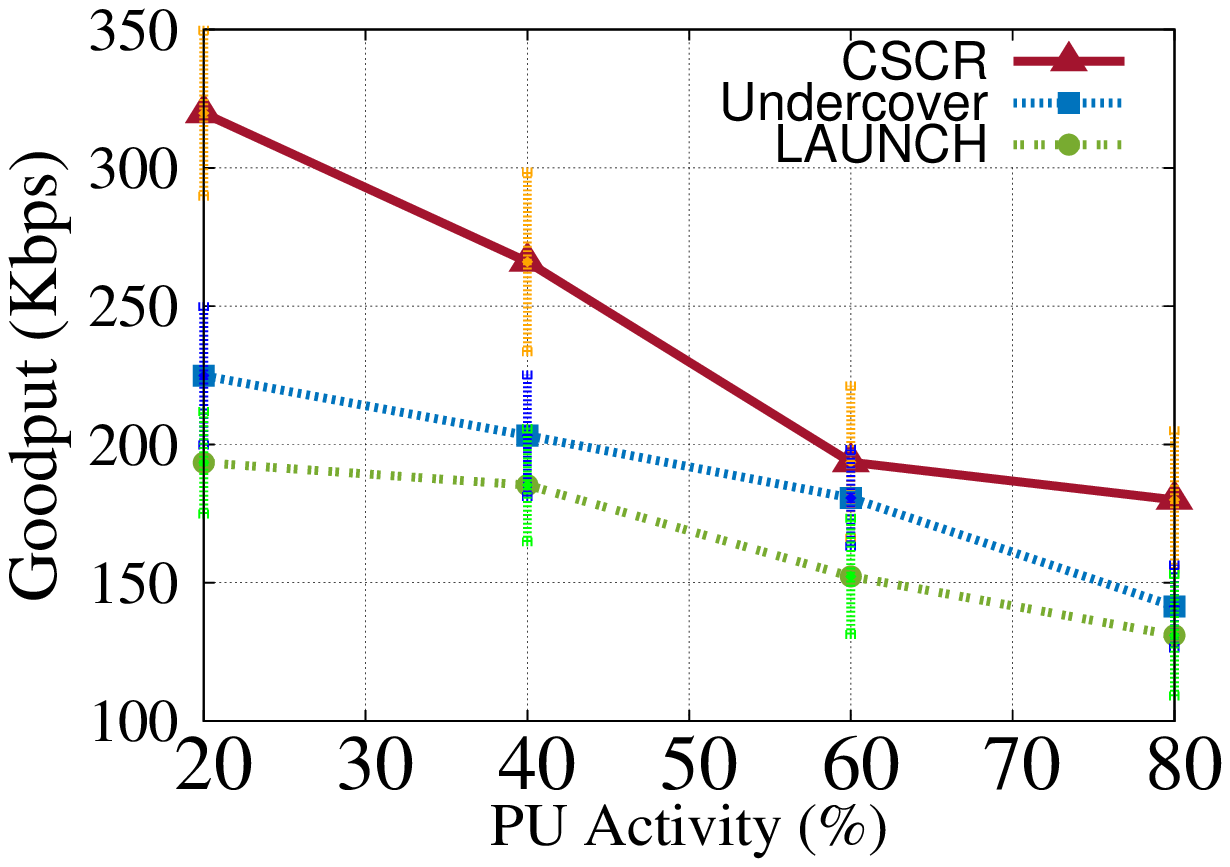}
	\caption{Goodput}
	\label{fig::puactivitythrough}
	\end{subfigure}
	\begin{subfigure}[t]{0.32\textwidth}
	\centering
    \includegraphics[width=2.2in]{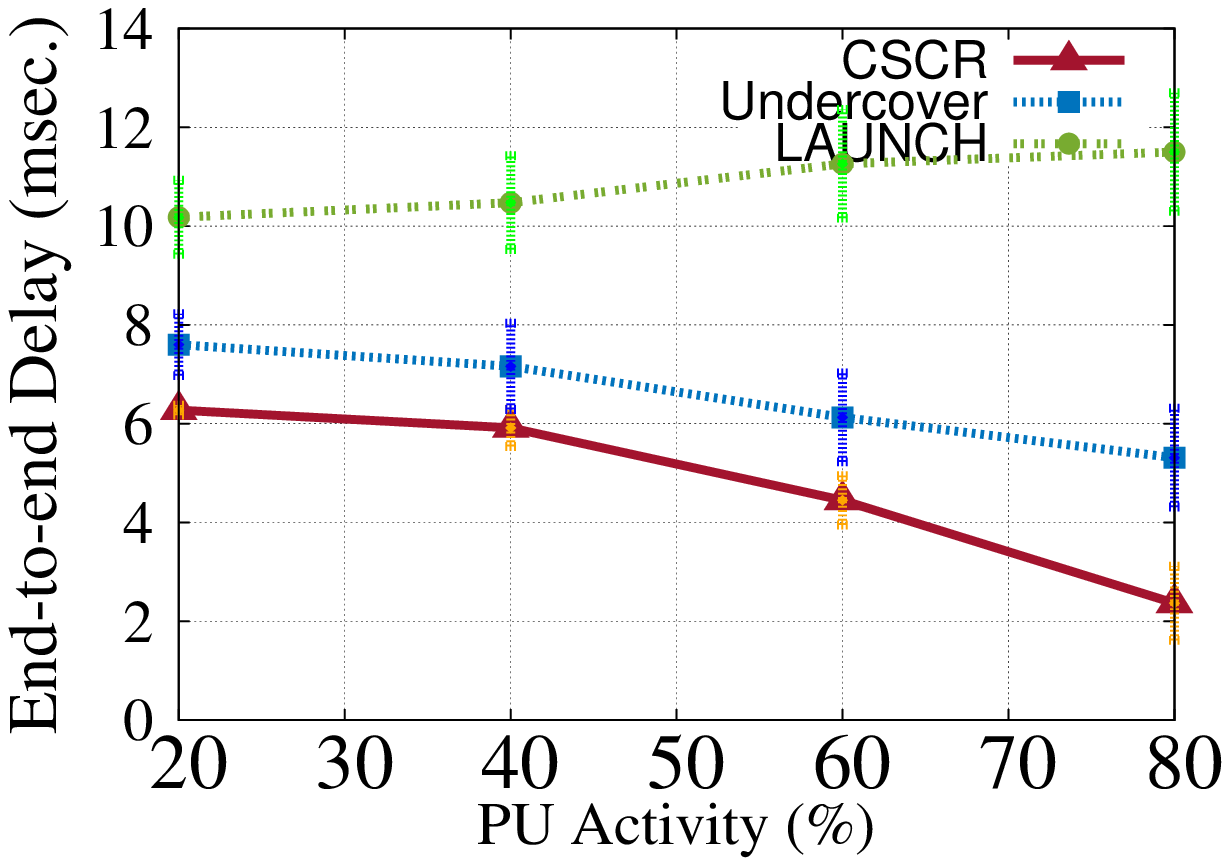}
	\caption{Average End-to-end Delay}
	\label{fig::puactivitydelay}
	\end{subfigure}
	\begin{subfigure}[t]{0.32\textwidth}
	\centering
    \includegraphics[width=2.2in]{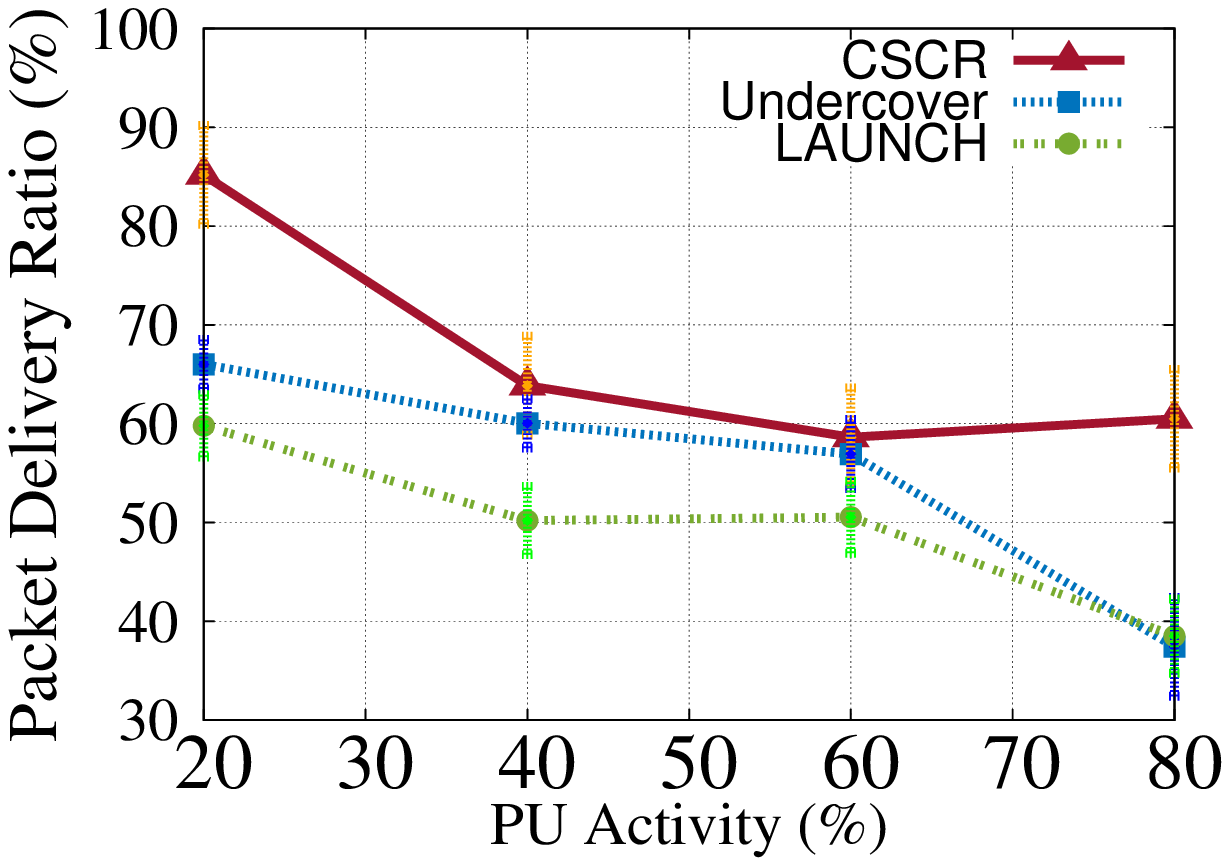}
	\caption{Packet Delivery Ratio}
	\label{fig::puactivitypdf}
	\end{subfigure}
\caption{Effect of changing PUs activity on network performance.}
\label{fig::puactivity}
\end{figure*}

\subsubsection{Changing PUs Activity}
Figure \ref{fig::puactivity} shows the effect of the PUs activity on the performance of the different routing protocols. We can note that the conclusions we have drawn from Figure \ref{fig::npus} can be also applied here. Based on that, we can conclude that the effect of increasing the PUs activity is quite similar to that of increasing the number of PUs.

\begin{figure*}[!t]
\centering
	\begin{subfigure}[t]{0.32\textwidth}
	\centering
    \includegraphics[width=2.2in]{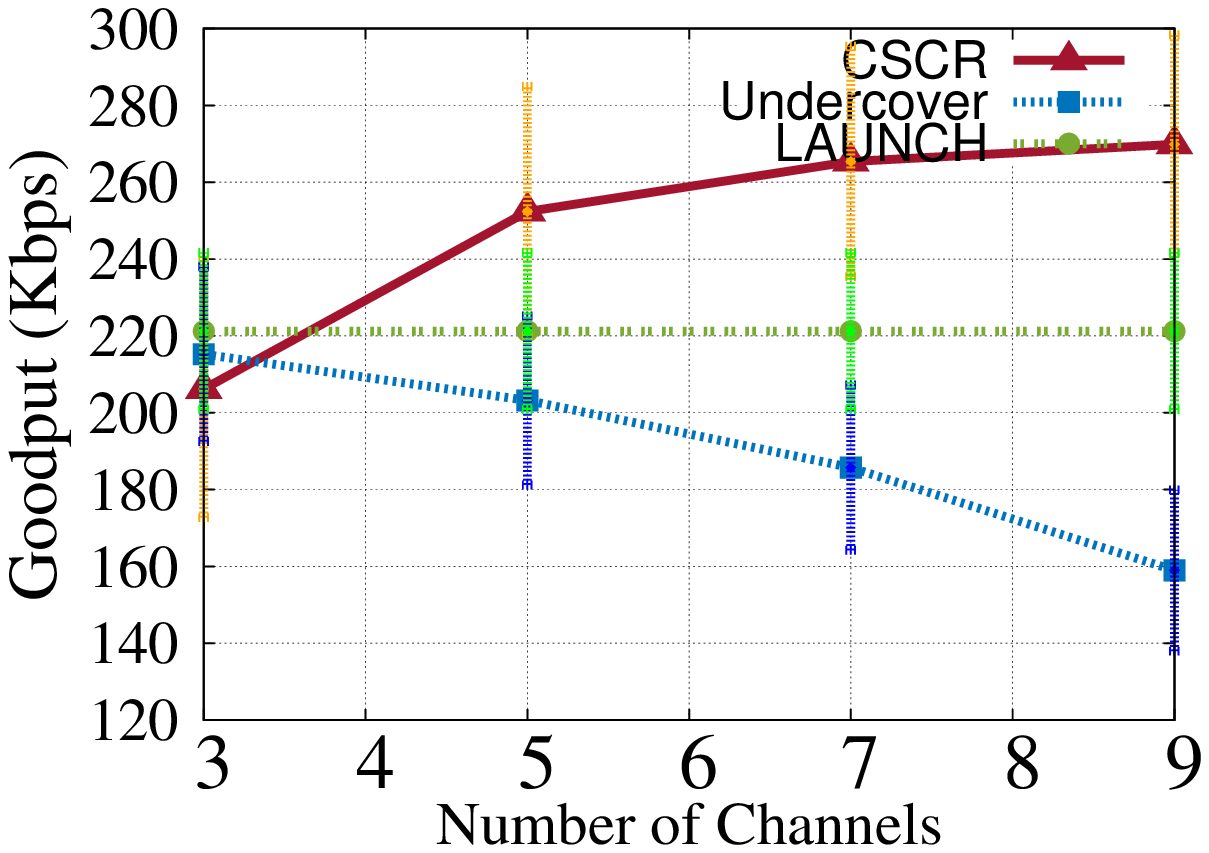}
	\caption{Goodput}
	\label{fig::nchannelsthrough}
	\end{subfigure}
	\begin{subfigure}[t]{0.32\textwidth}
	\centering
    \includegraphics[width=2.2in]{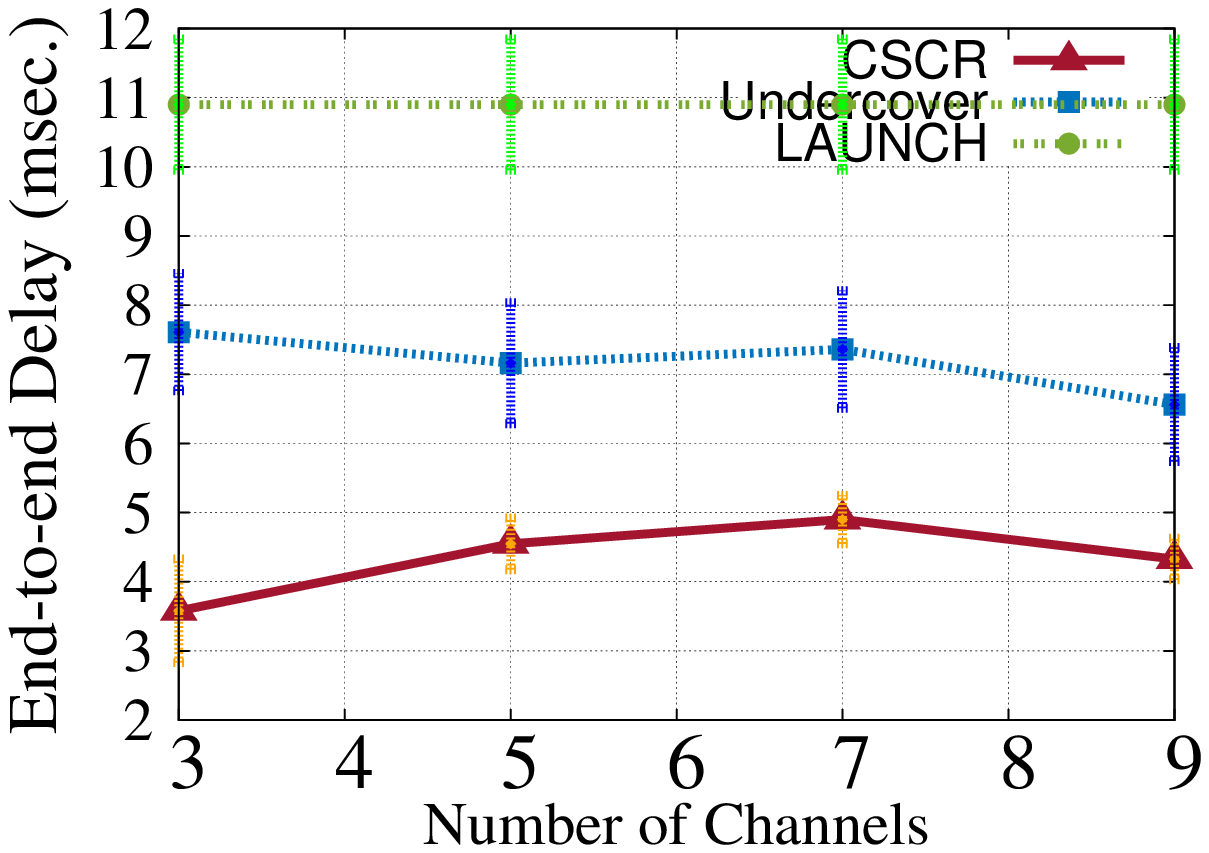}
	\caption{Average End-to-end Delay}
	\label{fig::nchannelsdelay}
	\end{subfigure}
	\begin{subfigure}[t]{0.32\textwidth}
	\centering
    \includegraphics[width=2.2in]{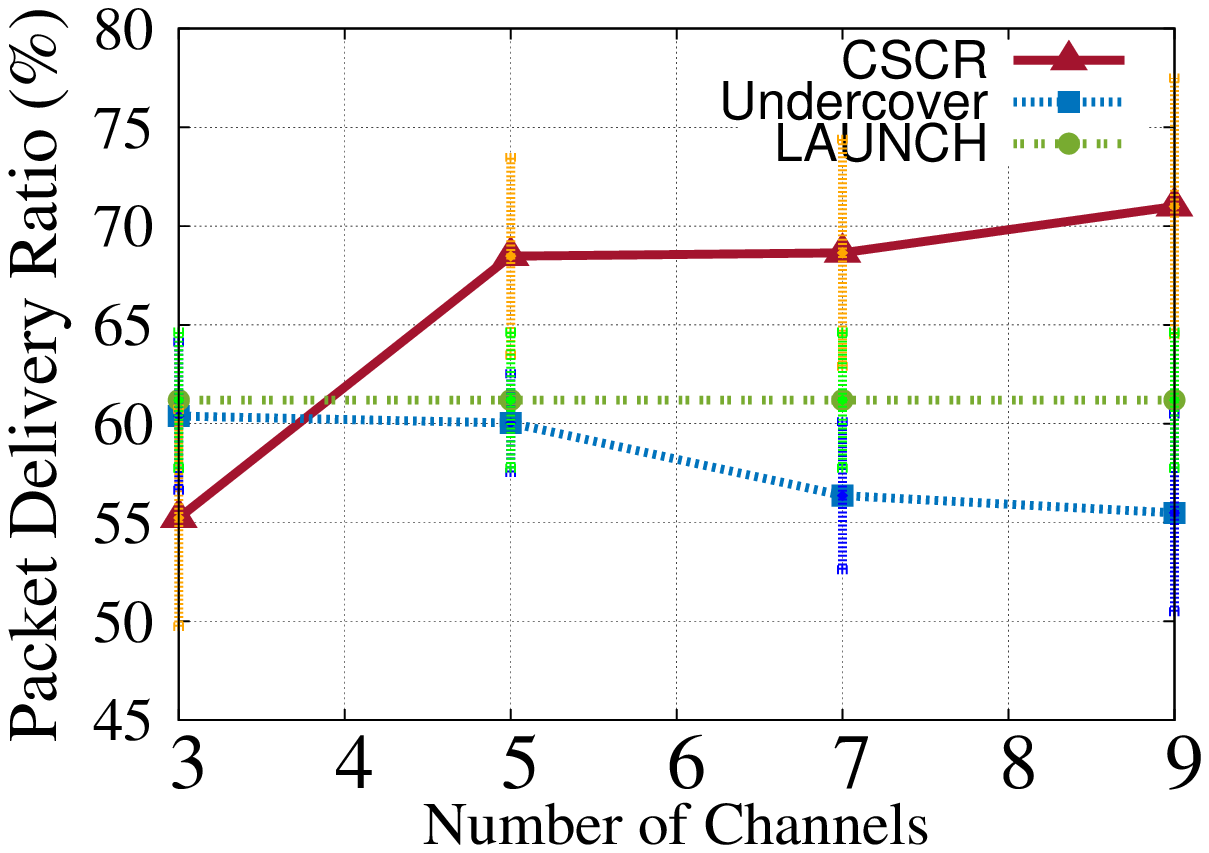}
	\caption{Packet Delivery Ratio}
	\label{fig::nchannelspdf}
	\end{subfigure}
\caption{Effect of changing number of Channels on network performance.}
\label{fig::nchannels}
\vspace{-0.35in}
\end{figure*}

\subsubsection{Changing Number of Available Channels}
Figure \ref{fig::nchannels} draws the effect of changing the number of channels on the performance metrics of the compared protocols. Studying this effect is important since we can evaluate the efficiency of our channel selection scheme through it. In addition, this figure draws some interesting results since, the three protocols behave completely different with changing the number of channels. For this reason, we discuss the behavior of each protocol separately. First, \sys{} performance is enhanced with the increase of the number of channels in terms of goodput (Figure \ref{fig::nchannelsthrough}) and the packet delivery ratio (Figure \ref{fig::nchannelspdf}). This happens since new opportunities and better routes are discovered when the number of available channels increases, giving \sys{} the ability to send data with a higher quality. The average end-to-end delay increases with the increase of goodput due to the increase of queues lengths and delays, which increases the number of transmissions as discussed before. The only exception happens at the end of the graph when the number of channels becomes higher than the number of PUs (at nine channels) as shown in Figure \ref{fig::nchannelsdelay}. At this point, \sys{} becomes able to send on - at least - one channel that is free of the existence of PUs. Thus the average end-to-end delay decreases. Generally, even in the worst case, when number of channels is minimal (Figure \ref{fig::nchannelsthrough} and \ref{fig::nchannelspdf}), the \sys{} performance is equal to\footnote{Although the average throughput and packet delivery ratio of \sys{} are lower than that of the other protocols, at three channels, as shown in these figures, they can be considered statistically equal since error bars of the three protocols overlap.} or better than the performance of the other protocols.

Second, although the Undercover strength is in its ability to construct cooperative groups, this technique becomes the reason of the performance degradation in this figure. Since Undercover does not know how to choose the best channel to send on, it may choose randomly a bad channel to use. Thus, we can see that the goodput, the packet delivery ratio and consequently the end-to-end delay, all decrease with the increase in the number of available channels.


Finally, LAUNCH shows an obscure performance with the increase in the number of channels. Although LAUNCH is considered a channel-aware protocol, it cannot make the perfect use of the newly introduced channels. This highlights an important fact in the design of LAUNCH. It is designed to minimize the switching delay as much as possible. This is done through choosing the channel which gives the least switching delay. In addition, a channel locking mechanism is used to hinder nodes from switching their sending channels freely. For this reason, source nodes always choose the same channels (which give the least delay) regardless of the number of the available channels, achieving the same performance in all cases.

\begin{figure*}[!t]
\centering
	\begin{subfigure}[t]{0.32\textwidth}
	\centering
    \includegraphics[width=2.2in]{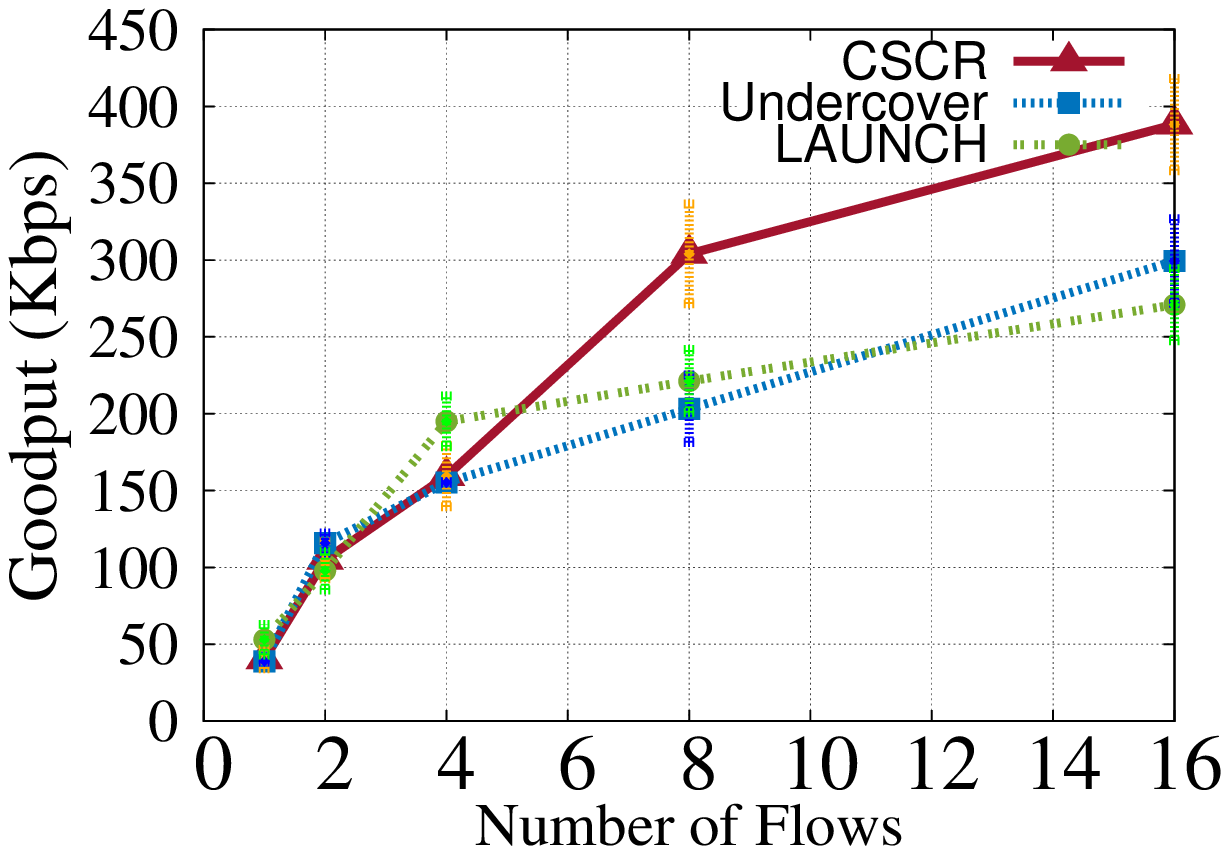}
	\caption{Goodput}
	\label{fig::nflowsthrough}
	\end{subfigure}
	\begin{subfigure}[t]{0.32\textwidth}
	\centering
    \includegraphics[width=2.2in]{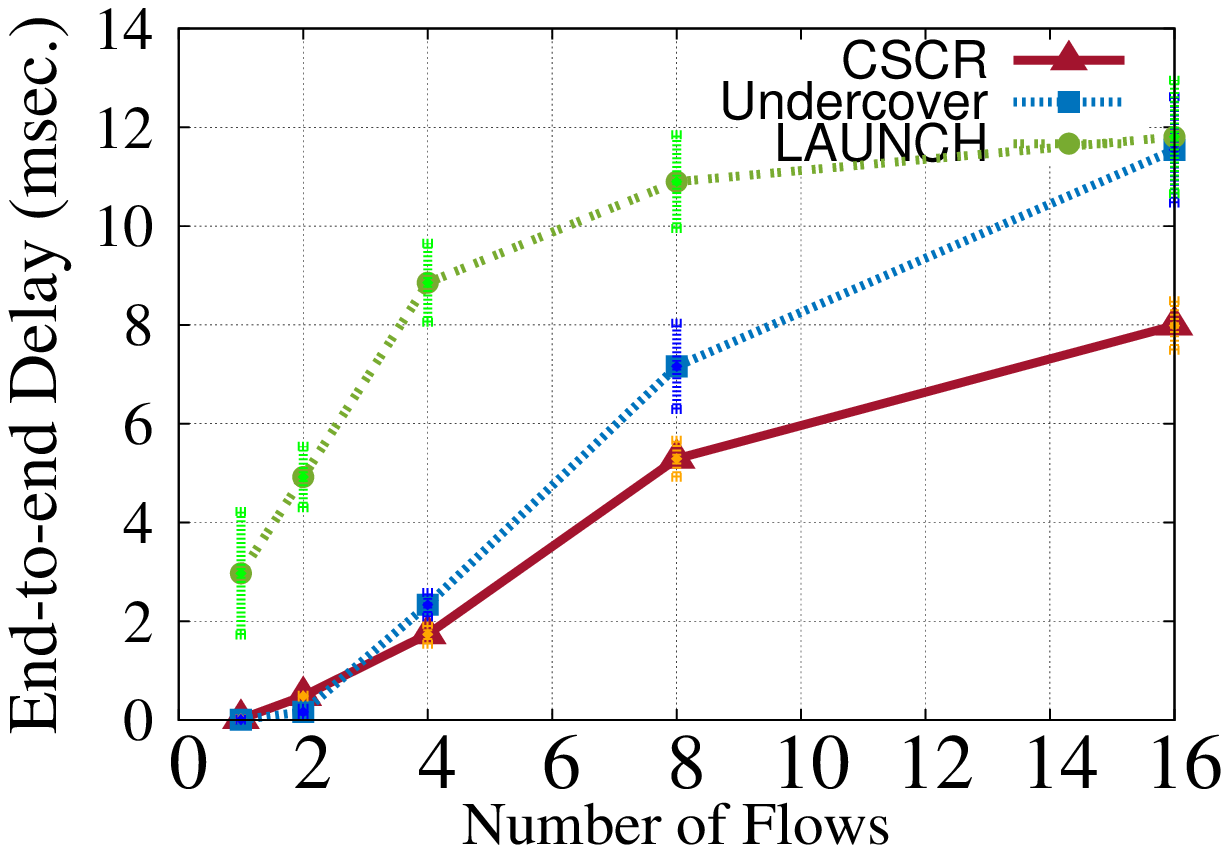}
	\caption{Average End-to-end Delay}
	\label{fig::nflowsdelay}
	\end{subfigure}
	\begin{subfigure}[t]{0.32\textwidth}
	\centering
    \includegraphics[width=2.2in]{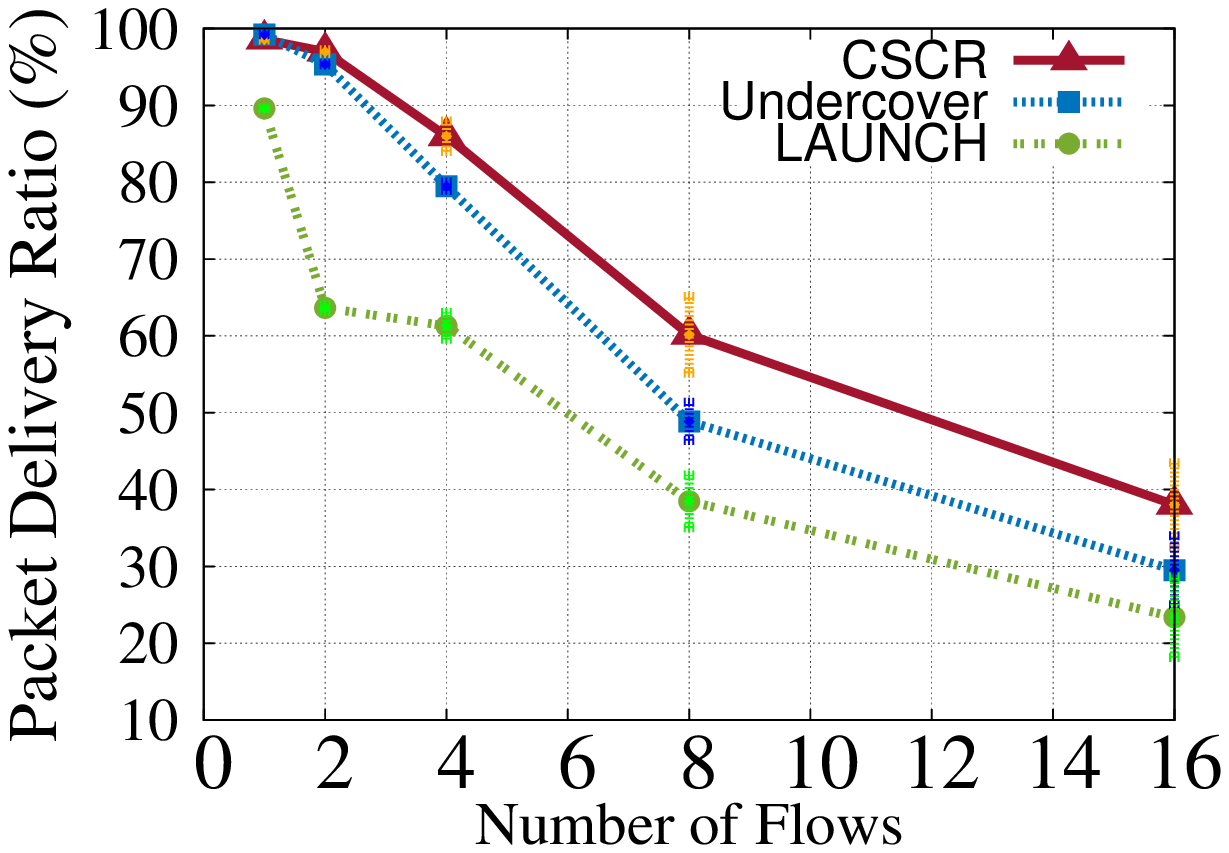}
	\caption{Packet Delivery Ratio}
	\label{fig::nflowspdf}
	\end{subfigure}
\caption{Effect of changing number of Flows on network performance.}
\label{fig::nflows}
\vspace{-0.35in}
\end{figure*}

\subsubsection{Changing Number of Flows}
Figure \ref{fig::nflows} shows the effect of changing the number of flows on the performance metrics. Increasing the number of flows increases the amount of data generated per second, increases the overall goodput (Figure \ref{fig::nflowsthrough}). Yet, this increase is accompanied by a decrease in the packet delivery ratio (Figure \ref{fig::nflowspdf}) due to the high interference between the SUs flows. As always, we can see that the behavior of the end-to-end delay is strongly related to that of the goodput as shown in Figure \ref{fig::nflowsdelay}. Also, conclusions we have got from the previous figures can also be applied to this one.

\section{Conclusion and Future Work} \label{sec::conc}
In this paper, we propose \sys{} which is a channel selection scheme for cooperative routing protocols in Cognitive Radio Networks. The cooperative protocol, we base our work on, utilizes cooperative communication techniques (cooperative diversity and cooperative beamforming) to enhance the quality of the signal at the receiver secondary user and null-out the transmissions at primary users. At each hop along the route, a group of nodes is chosen to send the data on some selected channel. The cooperative group and the operating channel are selected in a way that increases the achievable capacity, decreases the interference between secondary users flows, avoids the primary users activity areas, and decreases the channel switching delay. Simulations on NS2 are carried to evaluate the efficiency of \sys{} compared to two related protocols. Results show that \sys{} always outperforms both protocols in terms of network goodput, end-to-end delay, and the packet delivery ratio.

Future extensions of this work include investigating the effect of mobility on the cooperative routing and the channel selection. In addition, the scheme can be extended to support sending data to a multicast group instead of having one secondary receiver only. Finally, we can experiment the proposed scheme with other link layer contention management protocols.


\bibliographystyle{IEEEbib}
\bibliography{myLib}

\end{document}